\documentclass[aps,prl,reprint,groupaddress]{revtex4-2}
\usepackage{dcolumn,graphicx,physics,braket,hyperref}
\usepackage{color}
\usepackage{hyperref}
\usepackage{enumerate}
\usepackage{amsfonts}
\usepackage{amssymb}
\usepackage{comment}
\usepackage{mathtools,tikz}

\usepackage{amsthm}
\usepackage{bbm}
\usepackage{multirow}
\usepackage[colorinlistoftodos]{todonotes}
\usepackage[scr=boondoxo,frak=boondox,scrscaled=1.05]{mathalfa}
\usepackage{amsmath}

\DeclareMathOperator{\atantwo}{atan2}
\def\bra#1{\langle#1|}
\def\ket#1{|#1\rangle}

\def\cexp#1{\langle#1\rangle}

\def\tr#1{\mathrm{tr}\left(#1\right)}

\def\e{\mathrm{e}}

\def\ve{\varepsilon}

\renewcommand\vec{\boldsymbol}
\def\r{{\vec{r}}}

\def\p{{\vec{p}}}
\def\v{{\vec{v}}}

\def\A{{\vec{A}}}
\def\g{{\vec{g}}}
\def\u{{\vec{u}}}
\def\R{{\vec{R}}}
\def\G{{\vec{G}}}
\def\Q{{\vec{Q}}}

\begin{document}

\title{Dimensional reduction from magnetic field in moiré superlattices}
\author{Nisarga Paul}
\thanks{These authors contributed equally}
\author{Philip J.D. Crowley}
\thanks{These authors contributed equally}
\author{Liang Fu} 
\affiliation{Department of Physics, Massachusetts Institute of Technology, Cambridge, MA,
USA}

\begin{abstract}
    Moir\'e materials provide a highly tunable platform in which novel electronic phenomena can emerge. We study strained moiré materials in a uniform magnetic field and predict highly anisotropic electrical conductivity which switches easy-axis as magnetic field or strain is varied. The dramatic anisotropy reflects one-dimensional localization (dimensional reduction) of the electron wavefunctions along a crystal axis due to quantum interference effects. This can be understood in an effective one-dimensional quasiperiodic Aubry-André-like models, or in a complementary semiclassical picture. This phenomenon should be observable in strained moir\'e materials at realistic fields and low strain disorder, as well as unstrained systems with anisotropic Fermi surfaces.
\end{abstract}

\maketitle

\textbf{Introduction. } The advent of moir\'e materials has recently unlocked new opportunities for the control and engineering of two-dimensional quantum phases of matter \cite{Cao2018Apr,Andrei2020Dec,Mak2022Jul,Andrei2021Mar,Cao2018Apr2,Cao2020Jul,Park2021Feb,Tang2020Mar,Li2021Dec}. When subject to a  quantizing magnetic field, moir\'e systems exhibit a complex energy spectrum due to the interplay between Landau level physics and the moir\'e superlattice  \cite{Dean2013May,Ponomarenko2013May,Hunt2013Jun,Spanton2018Apr,Xie2021Dec,Kometter2022Dec}. 
Indeed, unlike ordinary semiconductors, a key feature of moir\'e systems is that the magnetic length and moir\'e lattice constant are often comparable $\ell_B \sim a_M$, both on the order of tens of nanometers. This allows access to two distinct sets of phenomena. First, the electronic spectrum exhibits fractal Hofstadter features, including Brown-Zak oscillations \cite{KrishnaKumar2018May,Yang2020Dec,Huber2022May}, due to the moir\'e unit cell enclosing $O(1)$ flux quanta \cite{Hofstadter1976Sep}, as observed in graphene/hBN \cite{Dean2013May,Ponomarenko2013May,Hunt2013Jun}. Second, the destructive interference between the Landau level orbitals and the moir\'e potential result in band flattenings, or magic zeros, at a discrete set of magnetic fields, and other commensurability phenomena \cite{Paul2022Sep,Weiss1989Jan}. \par

Strain is ubiquitous in realistic moir\'e materials. Since a small strain at the atomic scale is magnified by the large superlattice period, strain often plays an important role in understanding the phenomenology of the system\cite{Bi2019Jul,Mesple2021Sep,Shavit2023Feb,Zheng2021Oct,Nuckolls2023Feb}. Recent experimental advances in 2D materials also promise greater control over strain as a tuning knob, paving the way for ``strain engineering" \cite{Peng2020Nov} or ``straintronics" \cite{Miao2021Jun}.
The effect of strain on magnetotransport has just started to be studied experimentally and theoretically \cite{Vafek2022Sep,Wang2023Aug}. 

\begin{figure}[t!]
    \centering
\includegraphics[width=\linewidth]{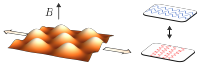}    \caption{\textbf{1D localization in strained moir\'e systems.} We consider strained moir\'e systems in a perpendicular field. The result is highly anisotropic conductivity which switches direction with $\rho$. This phenomenon is universal, accessible, and robust to strain disorder.}
    \label{fig:1}
\end{figure}

In this work, we show that strain provides an avenue to new physics in moir\'e superlattices: namely, the combination of strain (uniaxial or shear) and magnetic field generally induces one dimensional electronic states, leading to highly anisotropic magnetotransport (Fig. \ref{fig:1}). Moreover, the resistivity anisotropy, including the transport easy axis, is strongly tunable by carrier density $\rho$ or field. At fixed $\rho$, it alternates periodically in $1/B$ with period $\sim 2k_FQ$. This is due to an effective dimensional reduction of the 2D system to an array of 1D extended states. The origin of this dimensional reduction is the noncommutativity of projected position operators $x$ and $y$ in the effective picture of the Landau level perturbed by the moir\'e potential. 
We emphasize that this is a quantum interference phenomenon largely independent of the microscopic details of the moir\'e system. While highly anisotropic transport has been observed in moir\'e materials with a 1D superlattice structure \cite{Wang2022May,Kennes2020Feb,Beret2022Nov}, in our case the transport anisotropy varies periodically in $1/B$. Due to its universality and robustness, we expect field-induced dimensional reduction is a readily observable effect in moir\'e systems which is beyond the purview of traditional solids. \par 
\textbf{Dimensional reduction from strain and field. }First, let us review our expectations for magnetotransport in the 2D electron systems. In a clean system, when the Fermi level is between Landau levels, we expect familiar quantum Hall plateaus with vanishing $\sigma_{ii}$ ($i=x,y$) and $\sigma_{xy}$ quantized to $e^2\nu/h$. As the Fermi level sweeps through a Landau level, plateau transitions occur in $\sigma_{xy}$ and $\sigma_{ii}$ exhibits a peak. Our focus is on these peaks, and in particular oscillations of their heights. Specifically, we predict drastic anisotropy in the longitudinal conductivities in moir\'e systems due to the (generic) presence of strain.

\par 
To proceed, our starting point is the continuum Hamiltonian $(\hbar =1)$
\begin{equation}\label{eq:cont}
    H = H_0(\p-e\A)+ V(\r)
\end{equation}
which describes electrons minimally coupled to a uniform magnetic field in the presence of a moir\'e potential. We consider a moir\'e potential
\begin{equation}
    V(\r) = \sum_j V_j e^{i \vec Q_j \cdot \vec r}
\end{equation}
built out of long-wavelength harmonics with $Q_i \sim 1/a_M$. Let us specialize for now to a quadratic dispersion $H_0(\p) = p^2/2m$. When $V = 0$, the spectrum is described by Landau levels with energy $E_n = \omega_c(n+1/2)$, where $\omega_c = eB/m$. For small $V_0/\omega_c$, the Landau levels acquire a bandwidth of order $V_0$ and we can work within a low-energy effective theory. The effective Hamiltonian in the $n$th Landau level is given by $\tilde V = P_n VP_n$ where $P_n$ is the corresponding projector. Adopting a symmetric gauge, we find
\begin{equation}
    \tilde V(\tilde \r) = \sum_j \tilde V_j e^{i\vec Q_j \cdot \tilde{\vec r}}
\end{equation}
to leading order, where $\tilde V_j = V_j e^{-Q_j^2\ell^2/4}L_n(Q_j^2\ell^2/2)$ and $L_n$ is the $n$'th Laguerre polynomial. Here $\tilde{ \vec r} = (\tilde x,\tilde y)$ are the projected position operators, which satisfy 
\begin{equation}
    [\tilde x,\tilde y]=i\ell^2.
\end{equation}
\par
Consider first the example of a \textit{square} superlattice potential, $V(\r) = 2V_0 [\cos Q_x  x + \cos Q_y y]$. The projected potential is 
\begin{equation}
    \tilde V(\tilde \r) = 2 \tilde V_x \cos Q_x \tilde x + 2\tilde V_y  \cos Q_y \tilde y.
    \label{eq:Vtilde}
\end{equation}
Let $\beta = 2\pi/Q_xQ_y\ell^2$ be the magnetic flux per unit cell in units of $\hbar/e$. Then we may write
\begin{equation}\label{eq:AAH}
\tilde V = 2 \tilde V_y (\cos Y + \lambda \cos X),\qquad    [X,Y] = 2\pi i/\beta
\end{equation}
where $\lambda = \tilde V_x / \tilde V_y$. Let $X\ket{x} = x\ket{x}$ and let $c_n^\dagger$ create the state $\ket{\phi_n/Q_x}$ where $\phi_n = \phi_0 + 2\pi n/\beta$. For each $\phi_0$, we may regard $\tilde V$ as a 1D tight-binding model:
\begin{equation}
    \tilde V = \sum_n 2\tilde V_x \cos(\phi_n) c_n^\dagger c_n + \tilde V_y( c_{n+1}^\dagger c_n +\text{h.c.}).
\end{equation}
This is equivalent to the Aubry-Andr\'e-Harper (AAH) model \cite{AAH}, a canonical model which exhibits a localization-delocalization transition at $|\lambda|=1$. This is consistent with the duality of Eq. \eqref{eq:AAH} under $X\leftrightarrow Y$ and $\lambda \leftrightarrow 1/\lambda$. In the presence of strict $C_4$ rotational symmetry, $\lambda = 1$ and the model is at its critical point. However, any superlattice strain results in $Q_x\neq Q_y$ and can have stark effects, as $\lambda$, being the ratio of oscillating functions, fluctuates dramatically about the critical point. The appearance of the AAH model in the study of Bloch electrons in a magnetic field has been noted \cite{Rauh1974May}.

Previous studies have shown that this transition is sharp if and only if $\beta$ is a ``Diophantine number" \cite{BibEntry1999Nov}, which is a dense subset of the irrationals which excludes Liouville numbers (for which the wavefunctions never localize \cite{Avron1982Jan}). The existence of localized and delocalized phases at generic $\beta$, possibly separated by a mobility edge near $\lambda = 1$, has been verified numerically \cite{Modugno2009Mar}. \par 
Consider next the more realistic case of a \textit{triangular} superlattice cosine potential, $V(\r) = \sum_{j=1}^3 2V_j \cos \Q_j \cdot \r$, with $\Q_{j+1} = Q(\cos(2\pi j/3),\sin(2\pi j/3)$. The projected potential can be written 
\begin{equation}\label{eq:AAH3}
    \tilde V = \sum_{j=1}^3 2\tilde V_j \cos X_j,\qquad [X_j,X_{j+1}] = 2\pi i/\beta
\end{equation}
where $X_j = \vec Q_j\cdot \tilde \r$ and $\beta = 2\pi / |\Q_1 \times \Q_2|\ell^2$. For $\tilde V_2= \tilde V_3$, consider an $X_1$ eigenbasis $\ket{x_1}$ and let $c_n^\dagger$ create $\ket{\phi_n/Q_1}$ where $\phi_n = \phi_0 + 2\pi n /\beta$. Then for each $\phi_0$, we may again regard $\tilde V$ as a 1D tight-binding model:
\begin{equation}
    \tilde V = \sum_n 2\tilde V_1 \cos(\phi_n )c_n^\dagger c_n + 2\tilde V_2 \cos(\phi_{n+1/2}/2) (c_{n+1}^\dagger c_n + \text{h.c.}).
\end{equation}
In both of the examples Eq. \eqref{eq:AAH} and Eq. \eqref{eq:AAH3}, the presence of strain results in rapid switching between localized and delocalized regimes in the effective models with varying magnetic field. This implies a 1D localization in the 2D model along a direction which switches with varying magnetic field. For instance, in the square lattice case, for $|\tilde V_y| > |\tilde V_x|$ we expect wavefunctions localized along the $y$ direction, and for $|\tilde V_x| > |\tilde V_y|$, we expect wavefunctions localized along $x$. \par 

So far, we have discussed the square and triangular lattice cosine potentials in the perturbative regime. In fact, this perturbative result holds for any 2D periodic potential $V(\vec{r})$. In the absence of strain, $V(\vec{r})$ and $\tilde{V}(\tilde{\vec{r}})$  possess rotational symmetry. In the presence of strain, the eigenmodes of $\tilde V(\tilde \r)$ are generically delocalized in one direction 
and localized in the perpendicular direction. 
The localization direction 
may be simply read off from the functional form of $\tilde{V}(\tilde{\vec{r}})$: it is the direction in which $\tilde{V}(\tilde{\vec{r}})$ has extended level sets (i.e. open orbits). This direction is uniquely defined for generic $\tilde{V}(\tilde{\vec{r}})$, as the extended level sets of a 2D periodic function necessarily all run parallel to the same lattice vector $\vec{b}_\mathrm{ext.}$. As $\tilde{V}(\tilde{\vec{r}})$ is varied by tuning the strain, the direction of the extended level sets $\vec{b}_\mathrm{ext}$ can discretely switch along crystal directions~\cite{wilkinson1984critical,wilkinson1987exact,han1994critical,yeo2022non}~\footnote{It is straightforward to verify this argument for the square lattice potential whose extended levels sets run in the $x$ ($y$) direction for $\lambda>1$ ($\lambda < 1$) }. At the critical points where $\vec b_{\text{ext.}}$ changes (including in the zero-strain case), the wavefunctions are critically delocalized in all directions.
\par
While the discussion has so far dealt with the perturbative regime, we now demonstrate that this phenomenon is nonperturbative, and moreover holds for quite general energy dispersions. Our starting point is the semiclassical equations of motion for a Bloch wavepacket:
\begin{equation}\label{eq:pdot}
    \dot\p = -e  \dot \r \times \vec B,\quad 
    \dot \r  = \mathbf{\nabla} E(\p),
\end{equation}
where $E(\p)$ is the energy dispersion including the effect of the moir\'e potential. \par 
The relevant degrees of freedom at low temperatures are the electronic states near the Fermi surface. Wavepackets formed of these states can be thought of as propagating in a network made of copies of the \textit{original} Fermi surface separated by the superlattice wavevectors $\vec Q_j$. We will consider the case of a strained square moir\'e superlattice, $Q_x \neq Q_y$. Away from the junctions, electrons propagate freely and unidirectionally according to Eq. \eqref{eq:pdot} while picking up Aharanov-Bohm phases. Near the junctions, two incoming modes scatter into two outgoing modes, which is properly described as a Landau-Zener two-level crossing with scattering unitary
\begin{equation}\label{eq:U}
 U=   \begin{pmatrix}
    \sqrt{1-P}e^{-i\widetilde \varphi_S} & -\sqrt{P}\\
    \sqrt{P} & \sqrt{1-P}e^{i\widetilde \varphi_S}
    \end{pmatrix}
\end{equation}
where the $P$ is the magnetic breakdown probability:
\begin{equation}\label{eq:PLZ}
    P = e^{-2\pi/\delta}, \quad \delta = 16 eB v_1v_2 \sin\beta/E_{\text{gap}}^2.
\end{equation}
\begin{figure}
    \centering    \includegraphics[width=\columnwidth]{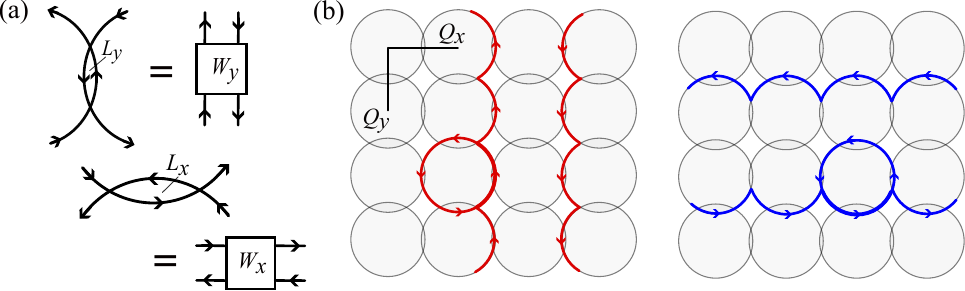}
\caption{\textbf{Semiclassical approach. } (a) $W_i$ denotes the scattering unitary across the lens orbit $L_i$ ($i=x,y$). (b) Fermi surfaces in the repeated zone scheme with moir\'e wavevectors $Q_x>Q_y$, with example semiclassical trajectories when electron motion is entirely in the $i$'th direction, which occurs at the quantization condition Eq. \eqref{eq:BS}.}
    \label{fig:semicl}
\end{figure}
Here $\v_1$ and $\v_2$ are the velocities of incoming electrons, $\hat{\vec v}_1 \cdot \hat{\vec v}_2=\cos \beta$, $E_\text{gap}=2V_0$ is the band gap at the Bragg plane due to moir\'e potential, and $\widetilde \varphi_S = \varphi_S-\pi/2$ with $\varphi_S = \pi/4-(\ln \delta +1)/\delta+\arg \Gamma(1-i/\delta)$  \cite{Paul2022Sep,Shevchenko2010Jul}. The form of $P$ makes clear that this approach is nonperturbative in $V_0/\omega_c$. We refer to \cite{PhysRevResearch.2.033271,Chambers1965Oct,Chambers1966Jul} for other examples of semiclassical network constructions. We remark that the regime of validity of the semiclassical approximation in this setting is large Landau level index.\par  
Due to the periodicity of the repeated Brillouin zone, energy eigenstates are Bloch-periodic eigenmodes of the network model. For this, it is instructive to study the scattering matrices $W_x$ and $W_y$ across the intersections of Fermi surfaces, which we call the \textit{lens orbits}, as indicated in Fig. \ref{fig:semicl}. These take the form \cite{Paul2022Sep}
\begin{subequations}\label{eq:W}
\begin{align}
    W_i&= \frac{1}{(1-P_i)e^{i(\xi_i+2\widetilde\varphi_{S,i})}-1} \begin{pmatrix}
    P_ie^{i\xi_i/2} & \kappa_i \\
    \kappa_i & P_ie^{i\xi_i/2}\\
    \end{pmatrix}\\
    \kappa_i &= e^{-i\widetilde\varphi_{S,i}}\sqrt{1-P_i}(e^{i(\xi_i+2\widetilde\varphi_{S,i})}-1)
\end{align}
\end{subequations}
where $\xi_i = \ell^2 S(L_i)$ and $S(L_i)$ is the $k$-space area of $L_i$. When $\kappa_i = 0$, $W_i$ is purely diagonal, implying that electron motion is entirely in the $i$'th direction ($i=x,y$). In other words, the junctions of $L_i$ are ``transparent" to the electrons. The condition for electron motion to be entirely in the $i$ direction is thus $\kappa_i = 0$ or\begin{equation}
\label{eq:BS}
\ell^2 S(L_i) + 2\widetilde{\varphi}_{S,i} = 2\pi n,
\end{equation}
where $n\in \mathbb{Z}$. This is a Bohr-Sommerfeld quantization condition for $L_i$.\par  

If $S(L_x) = S(L_y)$, then Eq. \eqref{eq:BS} is satisfied simultaneously for $x$ and $y$ if at all, and electron motion always remains delocalized in both directions. In the presence of any strain, however, $S(L_x) \neq S(L_y)$, so that electron motion can alternate between strict localization in $x$ and $y$, corroborating the perturbative approach. This approach places only topological constraints on the shape of the Fermi surface, and reveals that this phenomenon generalizes broadly and does not rely on perturbation theory.

\textbf{Magnetotransport. } To reveal the observable effects, we turn to a study of magnetotransport in strained moir\'e superlattices. Conductivity can be taken as $\sigma_{ab} = D_{ab}\tau$
under the assumption of a single relaxation time $\tau$, where $D_{ab}$ is the Drude weight \cite{Resta2018Sep}:

\begin{equation}\label{eq:Dab}
    D_{ab} = -ie \cexp{[j_a,x_b]}_0 + A\sum_{n\neq m} f_{nm} \frac{[j_a]_{nm} [j_b]_{mn}}{E_n-E_m}.
\end{equation}
Here, $j_a$ is the current density operator, $f_{nm} = f_n-f_m$ where $f_n = (1+e^{\beta(E_n-\mu)})^{-1}$ is the Fermi-Dirac distribution, and $A$ is the sample area. We derive this for completeness in the Supplementary Material. As an example, for a Fermi gas, $\vec j = -e\vec p/ mA$ and $\sigma_{ab} = (ie^2\tau/mA)\cexp{[p_a,x_b]}_0 = (ne^2\tau/m)\delta_{ab}$ is the familiar Drude conductivity. 
\par 
For a general Hamiltonian $H= T+V$ with $T = \sum_n E_n P_n$ (for projectors $P_n$) and $V$ small, for any operator $M$ we may consider the projected low-energy operator $\tilde M$ which satisfies $\cexp{\psi_1|M|\psi_2} = \cexp{\tilde \psi_1|\tilde M |\tilde \psi_2}$ for all energy eigenstates $\ket{\psi_i}$, and with $\ket{\tilde \psi_i} = P_n \ket{\psi_i}$. Expanding perturbatively in $V$, this is given by
\begin{equation}\label{eq:Mtilde}
    \tilde M = P_n M P_n + \sum_{m\neq n} \frac{P_n M P_m V P_n-P_n V P_m M P_n}{E_n-E_m} + O(V^2). 
\end{equation}
The low-energy effective current density operator is given to leading order by
\begin{subequations}
    \begin{align}
        \vec{\tilde{j}} &= ie [\tilde{\vec r}, \tilde V]/A\\
        &= e \ell^2   \sum_i   (\hat z\times \vec Q_i) \tilde V_ie^{i\vec Q_i\cdot \tilde r}/A
    \end{align}
\end{subequations}
by application of Eq. \eqref{eq:Mtilde}. Note that $\tilde j$ vanishes to zeroth order in $V$, as the bare current operator $\vec j$ only couples neighboring Landau levels. Finally, conductivity may be computed to leading order in $V/\omega_c$ using the effective operators $\tilde V, \tilde \r, \tilde{\vec j}$ in Eq. \eqref{eq:Dab}.
\par 

\begin{figure}
    \centering
\includegraphics[width=\columnwidth]{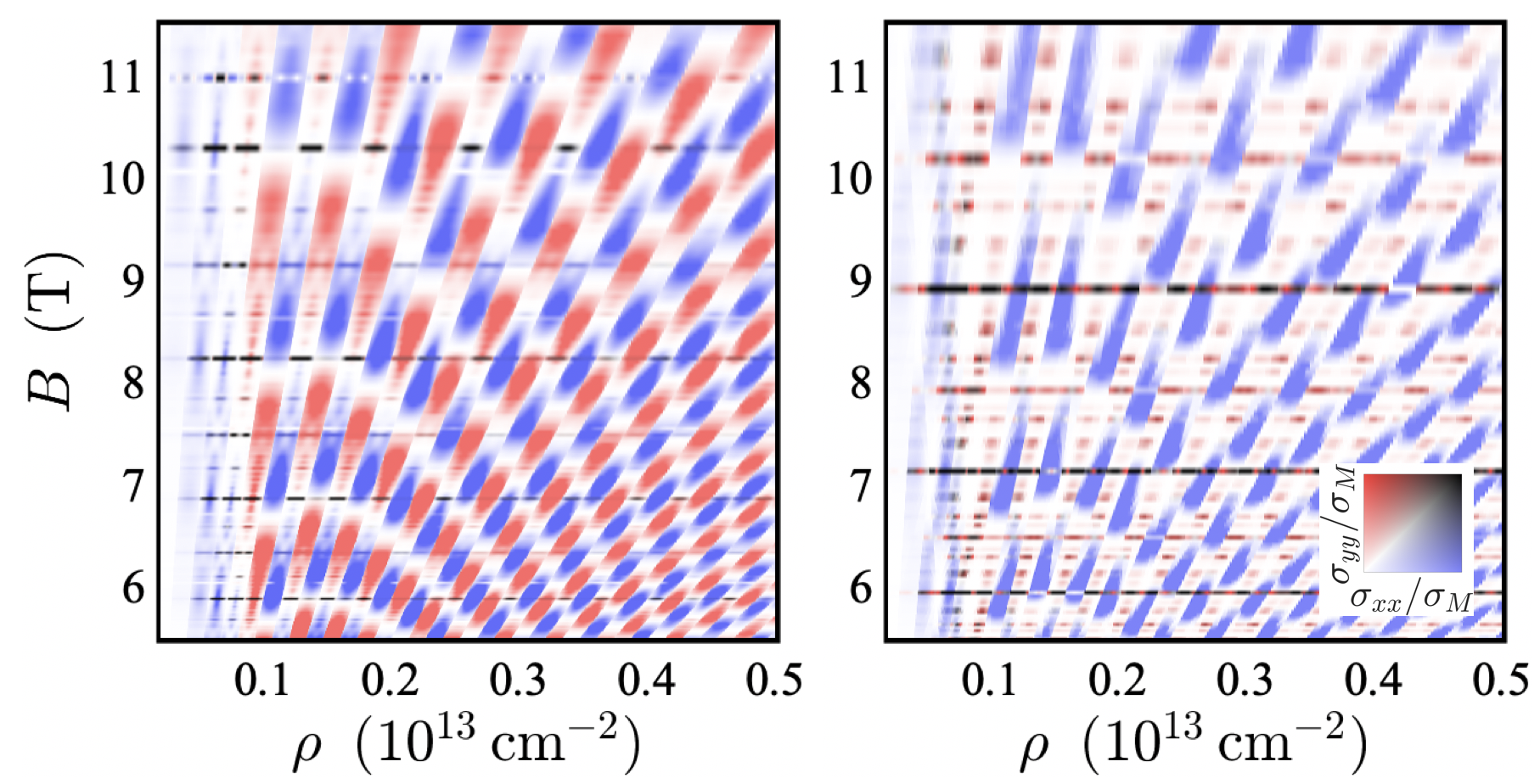}
\caption{\textbf{Conductivity switching.} $\sigma_{xx}$ and $\sigma_{yy}$ (normalized by some cutoff $\sigma_{M}$) in a strained square (left) and triangular (right) lattice and magnetic field, with uniaxial strain $x\mapsto \alpha x, y\mapsto \alpha^{-1} y$, $\alpha = 0.98$. We observe sharp conductivity anisotropy which switches as a function of $B$ /density. Horizontal dark features are Brown-Zak oscillations at flux $1/q$ per unit cell. ($Q=2\pi/10$ nm$^{-1}$, $V_0=$ 5 meV, $T = 0.5$ meV).}
\label{fig:strain}
\end{figure}

In Fig. \ref{fig:strain}, we use this method to plot longitudinal conductivities in the presence of a magnetic field and strain for square and triangular superlattices. Evidently  $\sigma_{xx}$ and $\sigma_{yy}$ show dramatic oscillations, consistent with our picture of localization-delocalization transitions, for the square superlattice, while $\sigma_{xx}$ does so for the triangular superlattice with milder oscillations in $\sigma_{yy}$. We also note the presence of Brown-Zak oscillations (horizontal features) at small-numerator rational flux values.\par 

We can infer the frequency of the directional switching as follows. In the presence of two unequal wavevectors $Q_1 \neq Q_2$, the switching is driven by the parameter $\lambda = \tilde V_1 / \tilde V_2$, which at large Landau level index $n$ behaves as
\begin{equation}\label{eq:omega0}
    \lambda \approx \sqrt{\frac{Q_2}{Q_1}} \frac{\cos(\sqrt{2n}Q_1\ell - \pi/4)}{\cos(\sqrt{2n}Q_2\ell-\pi/4)}
\end{equation}
where we used a large $n$ approximation to Laguerre polynomials \cite{szego1975}. The ratio of two incommensurate harmonics $\omega, \omega'$ generically exhibits fast and slow oscillations at frequencies $\omega+\omega'$ and $|\omega-\omega'|$. Relating density to Landau level index by $n = 2\pi \rho \ell^2$, we conclude that the fast oscillations are periodic in $1/B$ with frequency 
\begin{equation}
    \omega_0  = \sqrt{4\pi \rho} (Q_1 + Q_2)
\end{equation}
and slow modulating features are present at frequency $\sqrt{4\pi \rho}|Q_1-Q_2|$ (visible for larger strain). \par 
When $Q_1 \sim Q_2$, $\omega_0 \sim 2k_F Q$, which is the familiar frequency of Weiss oscillations in $1/B$ \cite{Paul2022Sep}. However, while Weiss oscillations exists even in a 1D potential, the phenomenon at hand requires a 2D potential, and thus it consolidates both Hofstadter and commensurability physics.\par 
\textbf{Discussion. } We have discussed a series of transitions observable as highly anisotropic and switching longitudinal conductivities, in the same universality class as the localization-delocalization transition of the AAH model, which can be readily observed in magnetotransport experiments on strained moir\'e materials. This is a manifestation of ``dimensional reduction" due to the fundamental quantum noncommutativity of position operators in the presence of a magnetic field.

In fact, this phenomenon can be observed more broadly in 2D materials with anisotropic Fermi surfaces and in a nonperturbative regime $V_0 \gtrsim \omega_c$, as we have demonstrated using a semiclassical approach. However, the magnetic length should be on the order of the lattice constant, making this difficult to realize outside moir\'e materials.  
\par 

Our predictions are remarkably robust to strain disorder, which is omnipresent in moir\'e systems. At large fields $(B \gtrsim \sqrt{\rho} Q)$, no reasonable strain disorder disrupts homogeneous 1D localization. At small fields $(B\ll \sqrt{\rho}Q)$ a strain disorder up to $\delta \alpha \sim \sqrt{ B/\sqrt{\rho}Q}$ is tolerable. For instance at $B=5$ T, $Q = 2\pi/10$ nm$^{-1}$ and one electron per unit cell, up to $30\%$ is permissible (far larger than the strain disorder in many clean moir\'e graphene samples). This is consistent with the fact that the switching frequency $\omega_0$ depends quite weakly on strain. \par 
We remark that evidence for unusual magnetotransport in moir\'e systems has been previously found in related contexts. Ref. \cite{Finney2022Apr} observed unusual magnetotransport in twisted bilayer graphene which was captured well by an anisotropic Hofstadter model. Such a model was also studied in Ref. \cite{Barelli1999Dec}, which noted enhanced $\sigma_{xx}/\sigma_{yy}$, but no ``switching". 
\par
One intriguing question for future work is whether this effect can be achieved in a general Chern band or in the absence of magnetic field. Indeed, the fundamental noncommutativity of the projected position operators depends only on the quantum geometry of the band. Other interesting directions for future studies include possible technological applications of this phenomenon, such as for ``moir\'e transistors" or magnetic sensors.

\par 
\begin{acknowledgments}
\textit{Acknowledgments---} We thank Trithep Devakul for collaboration on related work. We thank Caolan John and Patrick Ledwith for helpful discussions. This work is supported by the Air Force Office of Scientific Research (AFOSR) under Award No. FA9550-22-1-0432 and the Simons Investigator award from the Simons Foundation. 
\end{acknowledgments}

\bibliography{bib}

\widetext
\pagebreak

\begin{center}
\textbf{\large Supplemental Materials: Dimensional reduction from magnetic field in moiré superlattices}
\end{center}

\setcounter{equation}{0}
\setcounter{figure}{0}
\setcounter{table}{0}
\setcounter{page}{1}
\makeatletter
\renewcommand{\theequation}{S\arabic{equation}}
\renewcommand{\thefigure}{S\arabic{figure}}
\renewcommand{\bibnumfmt}[1]{[S#1]}
\renewcommand{\citenumfont}[1]{S#1}

\section{Strain in moir\'e bilayers}
\label{app:strain}
In this section we review, in general terms, the different kinds of strain that arise in moir\'e bilayers. In particular, we relate the heterostrain in the atomic layers to the strain in the moir\'e superlattice. 

\subsection{Strain tensor}

First, for a monolayer in which the atom at position $\r$ is subject to a displacement $\u(\r)$, we define the \textit{strain tensor} as
\begin{equation}
    \ve_{ij} = \frac12(\partial_i u_j + \partial_j u_i)
\end{equation}
and we refer to the diagonal components $\ve_{xx},\ve_{yy}$ as uniaxial strain and to $\ve_{xy}$ as shear strain. Next, consider a bilayer with strain tensors $\ve^{\ell}_{ij}$ where $\ell = \text{t},\text{b}$ for the top / bottom layer. We define the heterostrain and homostrain tensors as the layer-antisymmetric and symmetric combinations
\begin{equation}
\ve^{\text{het}}_{ij} = \ve^\text{t}_{ij}-\ve^\text{b}_{ij},\qquad \ve^{\text{homo}}_{ij} = \frac12(\ve^\text{t}_{ij}+\ve^\text{b}_{ij}).
\end{equation}
Both components are expected to exist for a generic strain configuration. 

\subsection{Effect on bilayer}

\textbf{Without strain. } Suppose we have a bilayer with primitive vectors $\R^\ell_j$, with $\ell = \text{t},\text{b}$ and $j=1,2$ which are related to a reference lattice with primitive vectors $\R_j$ via a twist angle $\theta_\ell$ and scale factor $\alpha_\ell$:
\begin{equation}
    \R^\ell_j = (1+\alpha_\ell)R(\theta_\ell)\R_j,\qquad R(\theta) = \begin{pmatrix}
        \cos\theta& -\sin\theta\\\sin\theta &\cos\theta
    \end{pmatrix}.
\end{equation}
If $\G_j$ are the reciprocal lattice vectors of the reference layer, the bilayer reciprocal lattice vectors are 
\begin{equation}
    \G^\ell_j = [(1+\alpha_\ell) R(\theta_\ell)]^{-T}\G_j
\end{equation}
and the moir\'e superlattice reciprocal vectors are, for small twist angles and scale factors,

\begin{equation}
    \g_j=\G_j^\text{t}-\G_j^\text{b} \approx -\underbrace{\begin{pmatrix}
         \alpha & \theta \\
         -\theta &  \alpha
    \end{pmatrix}}_{\equiv M(\alpha,\theta)}\G_j
\end{equation}
where $\alpha=\alpha_\text{t}-\alpha_\text{b}$ and $\theta = \theta_\text{t}-\theta_\text{b}$. The matrix $M(\alpha,\theta)$ effects a rotation by $\atantwo(-\theta,\alpha)$ and a scaling by $\sqrt{\alpha^2+\theta^2}$. 

\par 

\textbf{With strain. } Let us now consider the effect of strain on each layer given by the tensors $\ve^\ell_{ij}$. The corresponding action on the primitive vectors is 
\begin{equation}\label{eq:Rlj}
    \R_j^\ell = (I + \ve^\ell) \R_j.
\end{equation}
Including also the effects of twist and lattice mismatch, the reciprocal lattice vectors become
\begin{equation}
    \G^\ell_j = [(1+\alpha_\ell)R(\theta_\ell)(I+\ve^\ell)]^{-T}\G_j
\end{equation}
and the moir\'e superlattice vectors become
\begin{equation}
    \g_j'=\G_j^\text{t}-\G_j^\text{b} \approx -(M(\alpha,\theta) +\ve^{\text{het}}) \G_j
\end{equation}
for small $\alpha, \theta$ and $\ve^\ell$. The effect of homostrain drops out. Comparing $\g_j'$ (with strain) to $\g_j$ (without strain) we find that the consequence of heterostrain on the bilayer is an \textit{effective} strain on the moir\'e superlattice: 

\begin{equation}\label{eq:emoire}
    \g_j' = [I + \ve^{\text{moir\'e}}] \g_j,\qquad \ve^{\text{moir\'e}} \approx  \ve^{\text{het}}M(\alpha,\theta)^{-1}
\end{equation}
We are expressing the action of strain on the moir\'e superlattice wavevectors as if it were in real space (c.f. Eq. \eqref{eq:Rlj}) for convenience. Explicitly, 

\begin{equation}
    \ve^{\text{moir\'e}} = \frac{1}{\alpha^2+\theta^2} \begin{pmatrix}
        \alpha \ve^{\text{het}}_{xx}+\theta \ve^{\text{het}}_{xy} & \alpha \ve^{\text{het}}_{xy} -\theta \ve^{\text{het}}_{xx} \\ \alpha \ve^{\text{het}}_{xy} +\theta \ve^{\text{het}}_{yy} & \alpha \ve^{\text{het}}_{yy} -\theta \ve^{\text{het}}_{xy}  
    \end{pmatrix}
\end{equation}

We can take various limits of $\ve^{\text{moir\'e}}$. For example, in the presence of shear heterostrain and the absence of lattice mismatch, the moir\'e superlattice experiences \textit{uniaxial} strain proportional to $1/\theta$. In particular, for $\theta \sim 1^\circ$ the strain is \textit{magnified} by a factor of $\sim 50$, rendering it an important perturbation in realistic bilayers. Similar limits are summarized in Table \ref{table:1}.

\begin{table}
\centering
\begin{tabular}{|l |c| r|}
 \hline
 \textbf{bilayer heterostrain} & \,\,\,\textbf{bilayer distortion}\,\,\, & \,\,\textbf{moir\'e effective strain } \\ [0.5ex] 
 \hline
 uniaxial & lattice mismatch $\alpha$ & uniaxial $\sim 1/\alpha$\\
 uniaxial & twist $\theta$ & shear $\sim 1/\theta$\\
 shear & lattice mismatch $\alpha$ & shear $\sim 1/\alpha$\\
 shear & twist $\theta$ & uniaxial $\sim 1/\theta$\\
 \hline
\end{tabular}
\caption{Effect of atomic bilayer heterostrain on the moir\'e superlattice, expressed as an effective strain as in Eq. \eqref{eq:emoire}, in the presence of either twist or lattice mismatch. Homostrain has no effect. The effect on the moir\'e superlattice is magnified by a factor of $1/\theta$ or $1/\alpha$.}
\label{table:1}
\end{table}

\section{Effective Aubry-André models}
\begin{figure}[b!]
    \centering
\includegraphics[width=0.75\linewidth]{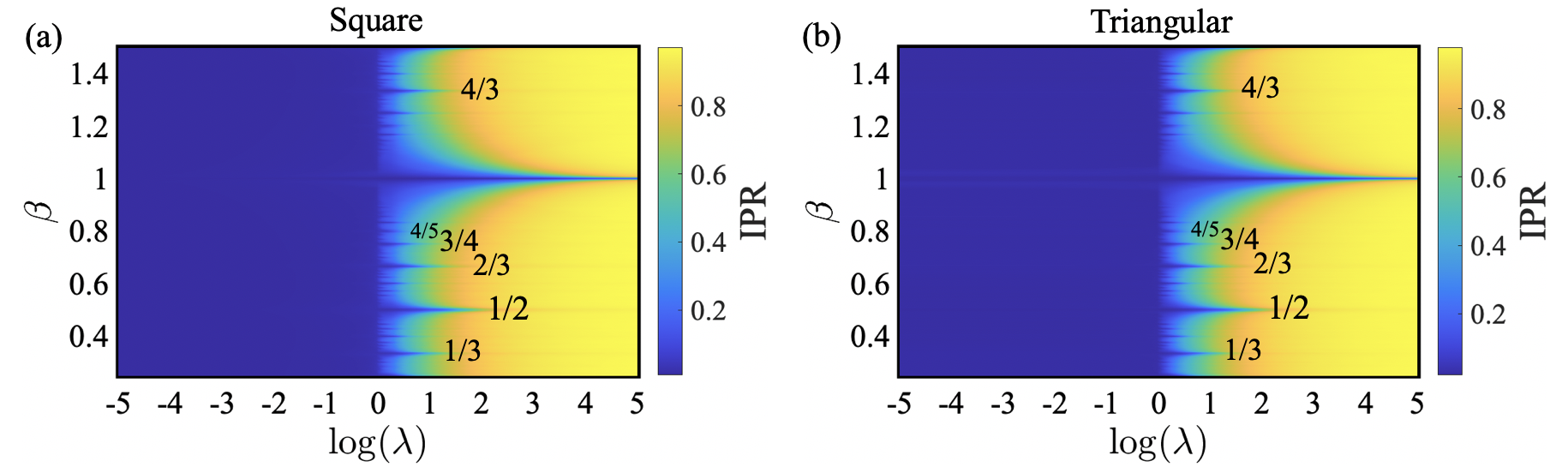}
    \caption{\textbf{Localization in Aubry-André models.} Inverse participation ratio $\sum |\psi_i|^4$ ($\sim$ inverse localization length) for models in Eq. \eqref{eq:AAs} on length-$N$ chains with $\beta = p/N$ for $p\in \mathbb{Z}$, averaged over $N_\phi$ values of $\phi_0$ ($N=173,N_\phi = 15$).}
    \label{fig:AAH3fold}
\end{figure}
In this section, we discuss the localization properties of the effective Aubry-Andr\'e models for the square and triangular superlattice cases. These correspond to 1D tight-binding models 
\begin{subequations}\label{eq:AAs}
    \begin{align}
        \tilde V_{\text{sq.}} &= \sum_n \lambda \cos(\phi_n) c_n^\dagger c_n + (c_{n+1}^\dagger c_n + \text{h.c.})\\
        \tilde V_{\text{tri.}} &= \sum_n \lambda \cos(\phi_n) c_n^\dagger c_n + \cos(\phi_{n+1/2}/2) (c_{n+1}^\dagger c_n + \text{h.c.})
    \end{align}
\end{subequations}
respectively, where $\phi_n = \phi_0 + 2\pi n/\beta$. When $\beta$ is a Diophantine number (which are dense in $\mathbb{R}$), the existence of a localization-delocalization transition has been rigorously established for $\tilde V_{\text{sq.}}$ \cite{BibEntry1999Nov}. \par

We plot the inverse participation ratio (IPR) $\sum_i |\psi(x_i)|^4$ averaged over $\phi_0$ of states in this model for varying $\lambda$ and $\beta$ in Fig. \ref{fig:AAH3fold}. These plots were generated taking $\beta$ rational, as is inevitable with numerics. Because the IPR is proportional to the inverse localization length in 1D systems when calculated in the thermodynamic limit, this supports the existence of a localization-delocalization crossover for generic $\beta$ near $\lambda =1$ (though not a strict phase transition). We remark that the IPR diagrams look quite similar.

\section{Landau level projection}
Here we review the method of Landau level projection used in the main text. We stress that this method is equivalent to evaluating the transport quantities to leading order in $V_0/\omega_c$. We consider the Hamiltonian of a 2D system lying in the $x,y$ plane
\begin{equation}
    H = H_0( \vec{\pi} ) + V ( \vec{r} ), \quad H_0 = \frac{p^2}{2m}, \quad V = \sum_{\vec{Q}} V_{\vec{Q}} \mathrm{e}^{ i \vec{Q} \cdot \vec{r}}
\end{equation}
where $\vec{\pi} = \vec{p} - e \vec{A}$, the background field is $\vec{A} = (B/2) (y,-x,0) $, and $V(\vec{r})$ is a weak periodic potential with reciprocal lattice vectors $\vec{Q}$ and Fourier coefficients satisfying $V_{\vec{Q}}^* = V_{-\vec{Q}}$.  We introduce the dual momenta $\vec{\tilde \pi} = \vec{p} + e \vec{A}$, and choose the symmetric gauge  so that $\vec{\pi}$ and $\vec{\tilde \pi}$ obey $[ \pi_x , \pi_y ] =  -i/\ell^2, [ \tilde \pi_x , \tilde \pi_y ] = i/\ell^2$ and $[\pi_i,\tilde \pi_j]=0$, where we've defined $\ell^2 = 1/eB$. We define the annihilation operators
\begin{equation}
    a = \frac{\ell}{\sqrt{2}}(\pi_x - i \pi_y), \quad b = \frac{\ell}{\sqrt{2}}(\tilde \pi_x + i \tilde \pi_y)
\end{equation}
which satisfy $[a,a^\dagger] = 1, [b,b^\dagger] = 1,$ and $[a,b]=0$. We define a basis satisfying $a^\dagger a\ket{n,m} = n\ket{n,m}, b^\dagger b\ket{n,m} = m\ket{n,m}$. Then $H_0 = \omega_c (a^\dagger a + \frac12)$ (where $\omega_c = eB/m$) and $V(\r)$ can be expressed in terms of the raising and lowering operators via

\begin{equation}
    \vec{r} = i\sigma_y \ell^2(\vec{\tilde{\pi}} - \vec{\pi}).
\end{equation}
 For small $V/ \omega_c$, the leading term in the effective Hamiltonian for the $n$'th Landau level is $\tilde V \equiv P_n V P_n$ (up to constant shift), where $P_n = \sum_m \ketbra{n,m}{n,m}$. This takes the form 
\begin{equation}\label{eq:generalVtilde}
    \tilde V = \sum_{\vec Q} \tilde V_{\vec Q} e^{i\vec Q \cdot \tilde{\vec r}}
\end{equation}
where $\tilde{\vec r} = (\tilde x,\tilde y)=i\sigma_y \ell^2 \vec{\tilde \pi}$, which satisfy $[\tilde x,\tilde y] = i\ell^2$, and $\tilde V_{\vec Q} = V_{\vec Q} D_{nn}((Q_x+iQ_y)\ell/\sqrt{2})$ with $D_{nn}(\eta) :=\bra{n} \exp \left[ i (\eta a + \eta^* a^\dagger ) \right] \ket{n} =  \e^{ - |\eta|^2/2}  L_n(|\eta|^2)$. $L_n(x)$ is the $n$'th Laguerre polynomial. \par 

To obtain the projected current density operator, we first review the general form of projected observables. For any Hamiltonian $H= T+V$ with $T = \sum_n E_n P_n$ and some operator $M$, we seek the projected operator $\tilde M$ which satisfies $\cexp{\psi_1|M|\psi_2} = \cexp{\tilde \psi_1|\tilde M |\tilde \psi_2}$ for all energy eigenstates $\ket{\psi_i}$, and with $\ket{\tilde \psi_i} = P_n \ket{\psi_i}$. Expanding perturbatively in $V$, this is given by
\begin{equation}
    \tilde M = P_n M P_n + \sum_{m\neq n} \frac{P_n M P_m V P_n-P_n V P_m M P_n}{E_n-E_m} + O(V^2). 
\end{equation}
In the case at hand, the current operator is
\begin{equation}
    \vec{J} = -\frac{\partial H}{\partial \vec{A}} = -\frac{e \vec{\pi}}{m}.
\end{equation}
We can write this in the form $\vec J = i e \omega_c \left( a^\dagger \vec{d}^* - a \vec{d} \right)$, which makes clear that $\vec J$ only couples neighboring Landau levels. We obtain
\begin{subequations}
    \begin{align}
         \vec{\tilde{J}} &= \frac{1}{\omega_c} P_n \vec J(P_{n-1}-P_{n+1})V P_n + \text{h.c.} \\
         &= -ie(P_n(a^\dagger \vec d^* - a \vec d)P_{n-1}V P_n - P_n(a^\dagger \vec d^* - a \vec d)P_{n+1}V P_n)+\text{h.c.}\\
         &=-ie (P_n a^\dagger \vec d^* V P_n + P_n a \vec d V P_n) + \text{h.c.}\\
         &= -ie P_n [\vec d a + \vec d^* a^\dagger,V]P_n\\
         &= ie P_n[ \tilde{\vec x}, V]P_n \\
        &= ie [\tilde{\vec x}, \tilde V]
    \end{align}
\end{subequations}

\section{Transport in perturbative regime}

Let us define and derive relevant quantities for transport. The most general form for conductivity can be written as \cite{Resta2018Sep,Scalapino1993Apr}
\begin{subequations}
    \begin{align}
        \sigma_{ab}(\omega) &= D_{ab}\left[ \delta(\omega) + \frac{i}{\pi \omega}\right] + \sigma^{\text{reg}}_{ab}(\omega)\\
        &= \sigma_{ab}^{\text{Drude}}(\omega)+\sigma^{\text{reg}}_{ab}(\omega)
    \end{align}
\end{subequations}
where $D_{ab}$ is the Drude weight and $\sigma^{\text{reg}}_{ab}(\omega)$ is the regular (or absorptive) part of the conductivity, sensitive to $\omega \neq 0$ transitions. Within a single relaxation time approximation with scattering time $\tau$, the DC conductivity is $\sigma_{ab} = D_{ab} \tau$. 
\par 
There are several methods to derive $D_{ab}$ \cite{Resta2018Sep,Scalapino1993Apr}, and we review one method here. In the absence of scattering the current response at short times to a DC electric field $E$ turned on a time $t=0$ is given by 
\begin{equation}
    j_a(t) = D_{ab} E_b t+O(E^2)
\end{equation}
where $j_a$ is the current density. Within the single relaxation time approximation the current response in an imperfect metal is given by truncating this at the scattering time $\tau$
\begin{equation}
    j_a(t) = D_{ab}E_b \tau + O(E^2)
\end{equation}
yielding a conductivity 
\begin{equation}
    \sigma_{ab} = D_{ab}\tau.
\end{equation}
This yields a straightforward way of calculating the Drude tensor at the level of linear response:
\begin{equation}
    D_{ab} = \lim_{t\to \infty} \left.\frac1t \frac{\partial\cexp{j_a}}{\partial E_b}\large\right|_{E=0}.
\end{equation}
Here $j_a$ is the current operator $j_a(t) = e^{iHt}j_ae^{-iHt}$ and $H$ is the Hamiltonian including an electric field, which we here include in Coulomb gauge
\begin{equation}
    H = H_0 -q \vec E\cdot \vec x
\end{equation}
with $\vec x$ the position operator. Differentiation yields 
\begin{equation}
    \left. \pdv{\cexp{j_a(t)}}{E_b}\large\right|_{E=0} = iq \int_0^tds\, \cexp{[j_a(t),x_b(s)]}_0
\end{equation}
Here we've defined $\cexp{O}_0=\tr{f_\beta(H_0-\mu)O}$ where $f_\beta(x) = 1/(1+e^{\beta x})$ is the Fermi-Dirac distribution. In Coulomb gauge the current density operator has only a paramagnetic term, and we may consequently use the relation 
$j_a = q\dot x_a / A$ where $A$ is the system area. Hence by integration $qx_a(t) - qx_a(s) = A \int_s^tds'\, j_a(s')$ and consequently 
\begin{equation}
\begin{aligned}
    q\int_0^tds\, x_a(s) 
    &= tqx_a(t) - A \int_0^tds\, \int_s^t ds'\, j_a(s')\\
    &= tqx_q(t) - A \int_0^t ds\, s j_a(s).
\end{aligned}
\end{equation}
Substituting back, we have 
\begin{equation}
    \begin{aligned}
        \left.\pdv{\cexp{j_a}}{E_b}\large\right|_{E=0} &= iqt\cexp{[j_a(t),x_b(t)]}_0 - iA \int_0^t ds\, s\cexp{[j_a(t),j_b(s)]}_0\\
        &= iqt\cexp{[j_a,x_b]}_0 - iA \int_0^t ds\, s\cexp{[j_a(t-s),j_b(0)]}_0.
    \end{aligned}
\end{equation}
Using $\lim_{t\to\infty} \frac1t \int_0^t ds\,s f(t-s) = \lim_{\eta\to 0^+} \int_0^\infty dt\,e^{-\eta t} f(t)$, we arrive at 
\begin{equation}\label{eq:Drude1}
    D_{ab} = iq \cexp{[j_a,x_b]}_0-i A\lim_{\eta\to 0^+} \int_0^\infty dt\, e^{-\eta t} \cexp{[j_a(t),j_b(0)]}_0.
\end{equation}
The second term can be rewritten in frequency space (leaving the $\eta\to 0^+$ implicit):
\begin{equation}
    \begin{aligned}
        -i A  \int_0^\infty dt\, e^{-\eta t} \cexp{[j_a(t),j_b(0)]}_0
        &= -i A  \int_0^{\infty}dt\, e^{-\eta t}\sum_{n} \frac{1}{1+e^{\beta(E_n-\mu)}} \cexp{n|e^{iE_n t}j_a e^{-iH_0t}j_b-j_b e^{iH_0 t}j_a e^{-iE_nt}|n}\\
        &=-i A  \int_0^{\infty}dt\, e^{-\eta t}(\sum_{n,m}\frac{e^{i(E_n-E_m)t}[j_a]_{nm}[j_b]_{mn}}{1+e^{\beta(E_n-\mu)}} - \sum_{n,m}\frac{e^{i(E_m-E_n)t}[j_b]_{nm}[j_a]_{mn}}{1+e^{\beta(E_n-\mu)}})\\
        &=  A \sum_{n,m} \left(\frac{1}{1+e^{\beta(E_n-\mu)}}-\frac{1}{1+e^{\beta(E_m-\mu)}}\right)\frac{[j_a]_{nm}[j_b]_{mn}}{E_n-E_m+i\eta}
    \end{aligned}
\end{equation}
yielding (for an electron with charge $q=-e$)
\begin{equation}\label{eq:Drude2}
    D_{ab} = -ie\cexp{[j_a,x_b]}_0 + A \sum_{n\neq m} f_{nm} \frac{[j_a]_{nm} [j_b]_{mn}}{E_n-E_m}
\end{equation}
where $f_{nm} = f_n-f_m$ and $f_n = (e^{\beta(E_n-\mu)}+1)^{-1}$. Note this obtains the correct Drude weight for a Fermi gas, in which case $\vec j = -e\vec p/mA$ and
\begin{equation}
    D_{ab} = \frac{ie^2}{mA}\cexp{[p_a,x_b]}_0 = \frac{ne^2}{m} \delta_{ab}\tag{Fermi gas}
\end{equation}
where we've used $\cexp{[p_a,x_b]}_0 = -iN \delta_{ab}$ and $[j_a,j_b]=[j_a,H_0]=0$. We also obtain the correct (vanishing) Drude weight for a 2D electron gas in a magnetic field. The first term yields an identical contribution $(ne^2/m)\delta_{ab}$, while the second term can be simplified using
\begin{equation}
    [j_a]_{cq;dq'} = e\delta_{qq'} (\sqrt{c+1}\delta_{c+1,d} + s \sqrt{c} \delta_{c-1,d})/(\sqrt{2s} \ell m A)
\end{equation}
where $s=\pm 1$ for $a=x,y$ and $q$ is the intra-Landau level quantum number (e.g. angular momentum if symmetric gauge is chosen). We then obtain
\begin{equation}
    \begin{aligned}
        D_{aa} &= \frac{ne^2}{m} + \sum_{c\neq d,q,q'} f_{cd} \frac{[j_a]_{cq;dq'} [j_b]_{dq';cq}}{E_c-E_d} \\
        &= \frac{ne^2}{m} +\frac{e^2}{2s\ell^2m^2A} \sum_{c\neq d,q,q'} f_{cd} \delta_{qq'}\frac{s d \delta_{c+1,d}+sc\delta_{c-1,d}}{E_c-E_d}\\
        &= \frac{ne^2}{m} +\frac{e^2}{2\ell^2m^2A} \left(\sum_{d,q}f_{d-1,d}\frac{d}{-\omega_c} + \sum_{c,q} f_{c,c-1}\frac{c}{\omega_c} \right)\\
        &= \frac{ne^2}{m} - \frac{ne^2}{m} = 0.
    \end{aligned}\tag{Uniform B field}
\end{equation}
This is the familiar fact that longitudinal conductivity vanishes for Landau levels at any chemical potential or temperature. Indeed, this is a consequence of the $f$-sum rule and Kohn's theorem \cite{Kohn1961Aug}, which dictates that the total optical absorption in Landau levels is saturated by the absorption peak at $\omega=\omega_c$.
\par 

Applying Eq. \eqref{eq:Drude2} to the model studied in the main text we make the replacements $H_0 \to \tilde{V}$, $j_a \to i e [\tilde{x}_a ,\tilde{V}]/A$, $x_a \to \tilde{x}_a$, $q \to - e$ where $\tilde V, \tilde x_a$ are the projected potentials and position operators, respectively. This projected method is ideally suited to our calculations because the Drude weight is a DC response which is not sensitive to matrix elements in states outside the low-energy subspace. 

\textbf{Numerics. } Our numerical calculations are done for each $Q_x Q_y\ell^2/2\pi \equiv \beta^{-1} = p/q$ on a finite system of $\sim N\times q$ unit cells (thus enclosing $N \times p$ flux quanta and maintaining moir\'e translation symmetry) for $q \in N_q\mathbb{Z}$. We found $N\sim 30, N_q \sim 50$ more than sufficient for convergence.

\section{Shear strain}

We framed the discussion in the main text around uniaxial heterostrain. In this section, for completeness, we show that shear heterostrain leads to unchanged main results. Let's consider a shear strain $\alpha$ on a square moir\'e superlattice acting as 
\begin{equation}
    \begin{pmatrix}
        \Q_x'\\ \Q_y'
    \end{pmatrix}  = \begin{pmatrix}
        1 & \alpha \\ 0 & 1
    \end{pmatrix}\begin{pmatrix}
        \Q_x\\ \Q_y
    \end{pmatrix}.
\end{equation}
Here $\Q_x = Q\hat x$, $\Q_y = Q\hat y$. Let's define $Q' = \sqrt{1+\alpha^2} Q$. The moir\'e potential is $V(\r) = 2V_0[\cos \Q_x'\cdot \r + \cos \Q_y'\cdot \r].$
The potential projected into the $n$'th Landau level is 
\begin{equation}
    \tilde V(\tilde \r) = 2 \tilde V_x \cos \Q_x' \cdot \tilde \r + 2 \tilde V_y \cos \Q_y' \cdot \tilde \r
\end{equation}
where $\tilde \r = (\tilde x,\tilde y)$, $[\tilde x,\tilde y] = i\ell^2$, $\tilde V_x = V_0 e^{-Q'\ell^2/4} L_n(Q'^2\ell^2/2)$, and $\tilde V_y = V_0 e^{-Q^2\ell^2/4} L_n(Q^2\ell^2/2)$. It can be written 
\begin{subequations}
\begin{align}
        \tilde V &= 2 \tilde V_x (\cos X + \lambda \cos Y)\\
        [X,Y] &= i Q^2\ell^2,\quad \lambda = \tilde V_y / \tilde V_x
\end{align}
\end{subequations}
This again is an AAH model. Eigenmodes are generically localized in the diagonal basis of $X$ or $Y$ depending on whether $|\lambda|<1$ or $|\lambda| > 1$.

\end{document}